\documentclass[12pt]{iopart}

\usepackage{iopams}
\usepackage{color}
\usepackage{graphicx}
\usepackage{comment}
\usepackage{cite}
\usepackage{url}
\usepackage{doi}
\usepackage{hyperref}

\renewcommand{\bs}{\boldsymbol}

\newcommand{\OK}{}

\begin{document}

\title{How tracer particles sample the complexity of turbulence}

\author{Cristian C. Lalescu$^{1}$, Michael Wilczek$^{1}$}
\ead{michael.wilczek@ds.mpg.de}
\address{$^1$ Max Planck Institute for Dynamics and Self-Organization, Am Fassberg 17, D-37077 G\"ottingen, Germany}

\date{\today}

\begin{abstract}

On their roller coaster ride through turbulence, tracer particles sample the fluctuations of the underlying fields in space and time. Quantitatively relating particle and field statistics remains a fundamental challenge in a large variety of turbulent flows. We quantify how tracer particles sample turbulence by expressing their temporal velocity fluctuations in terms of an effective probabilistic sampling of spatial velocity field fluctuations. To corroborate our theory, we investigate an extensive suite of direct numerical simulations of hydrodynamic turbulence covering a Taylor-scale Reynolds number range from 150 to 430. Our approach allows the assessment of particle statistics from the knowledge of flow field statistics only, therefore opening avenues to a new generation of models for transport in complex flows.

\end{abstract}

\section{Introduction}

Turbulence, as frequently encountered in our atmosphere and oceans as well as in a plethora of astrophysical phenomena, exhibits a stunning complexity in space and time. This complexity can be perceived from two complementary perspectives: from the Eulerian point of view turbulence is observed in the laboratory frame by probing the flow at one or more fixed spatial points; in the Lagrangian frame, the flow is probed by tracer particles which follow the velocity field (see movie SM1 which is available online at \url{http://stacks.iop.org/NJP/20/013001/mmedia}). This perspective is of particular importance for characterizing turbulent mixing and transport in particle-laden flows. 

One of the most challenging phenomena of turbulence is the breaking of statistical self-similarity: probing velocity fluctuations across scales reveals a continuous change in shape of the corresponding distributions from the largest, energy-containing scales of the flow down to the smallest, dissipative scales, on which extreme events occur with astonishing frequency \cite{sreenivasan97arf,ishihara07jfm}. This phenomenon, known as intermittency, can be cleanly isolated in homogeneous isotropic turbulence, which shares universal small-scale properties with a wide class of hydrodynamic flows even at moderate Reynolds numbers \cite{schumacher14pnas}. Experimental \cite{douady91prl,cadot95pof,ganapathisubramani08jfm,elsinga10jfm,worth11pre} and numerical \cite{ashurst87pof,she90nat,vincent91jfm,jimenez93jfm,vincent94jfm,moisy04jfm,leung12jfm,yeung15pnas} studies of isotropic turbulence have traced the signatures of Eulerian statistics back to the intricate spatial structure of turbulence. As tracer particles sample the fine-scale structure of turbulence, they encounter violent velocity and acceleration fluctuations on the smallest temporal scales \cite{mordant01prl,laporta01nat,mordant04phd,biferale06jot}. In analogy to the Eulerian observations, Lagrangian velocity increment distributions are also non-self-similar \cite{mordant01prl} when probing the velocity fluctuations across temporal scales. However somewhat puzzlingly, extreme events are even more frequent in the Lagrangian than in the Eulerian frame.

This leads to the long-standing riddle of relating the observations in the two complementary frames, and in particular to the question:
how do tracer particles sample the spatio-temporal velocity fluctuations of turbulence?
Here we provide an answer by combining an exact probabilistic framework first introduced in~\cite{kamps09pre,homann09njp} with the physics of Lagrangian particle transport and Eulerian temporal decorrelation into a predictive theory relating Lagrangian to Eulerian statistics. We show that, remarkably, \emph{temporal} velocity fluctuations along Lagrangian tracers can be statistically predicted from \emph{instantaneous spatial} velocity fluctuations by properly mixing Eulerian statistics of various scales. We demonstrate that this probabilistic mixing can be captured in terms of the probability density function (PDF) of an effective Lagrangian dispersion process, whose properties we determine. The investigation of an extensive suite of direct numerical simulations of homogeneous isotropic turbulence furthermore reveals its universality with respect to a range of Reynolds numbers. 

\section{Bridging Eulerian and Lagrangian statistics}

\begin{figure}
    \begin{center}
    \includegraphics{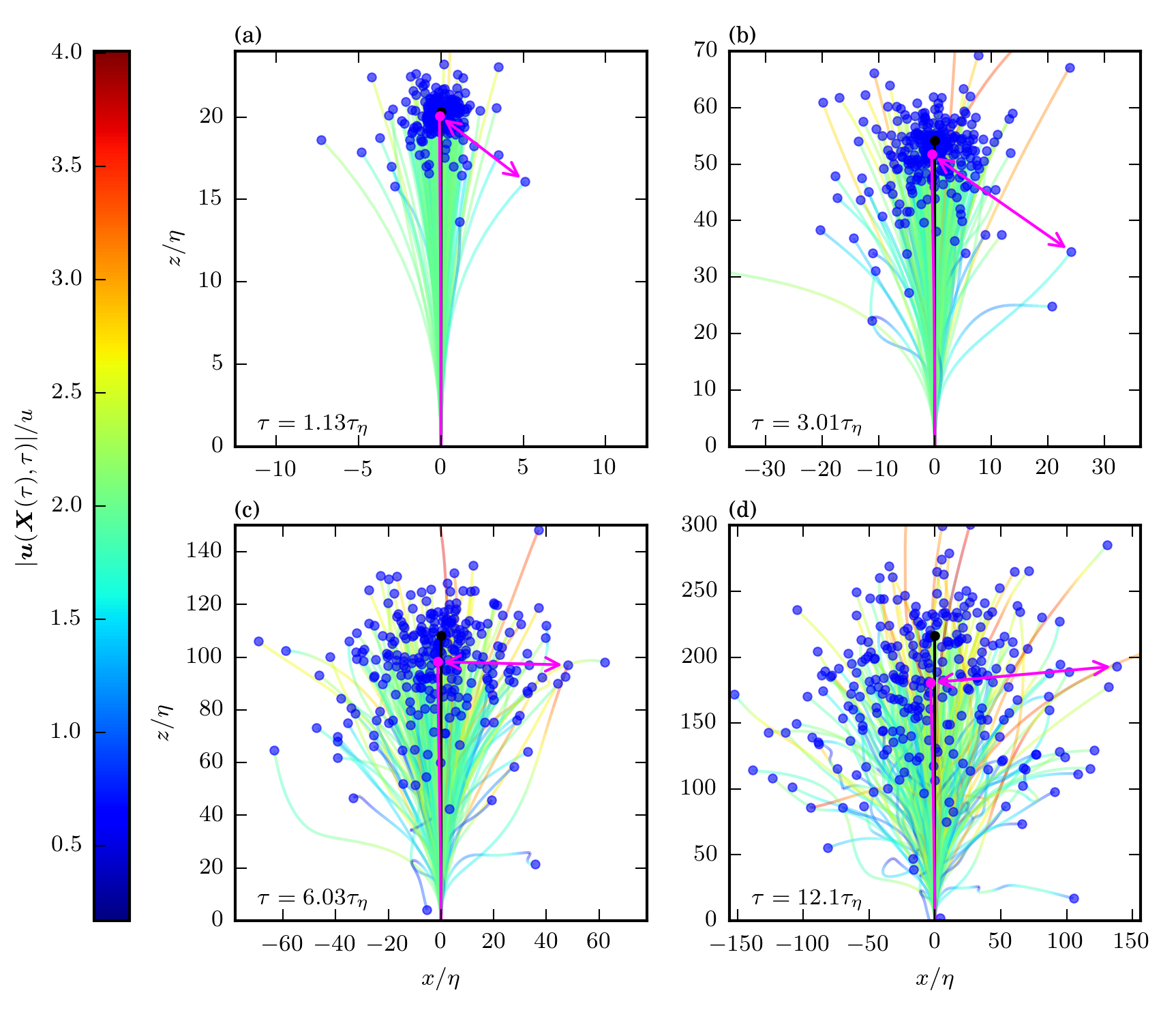}
    \end{center}
\caption{
Dispersion of Lagrangian trajectories with fixed initial velocities (twice the rms velocity $u$) from a direct numerical simulation at $R_\lambda = 316$.
For this visualization, particles with identical initial velocity magnitudes have been selected from the flow domain and shifted to a common origin and orientation. (a): Initially, their motion is almost ballistic and remains close to the mean Lagrangian trajectory shown in magenta.
(b)-(d): Over time, the particles disperse and randomly sample the region surrounding the mean trajectory.
The black curve shows the ballistic approximation to the mean Lagrangian trajectory.
    The distance between the mean Lagrangian trajectory and the individual trajectories (as indicated for one sample trajectory by the arrow) is the effective Lagrangian displacement. (See movie SM2 for an animated version of this figure which can be found online at \url{http://stacks.iop.org/NJP/00/000000/mmedia}.)}
\label{fig:effectivelagrangiandispersion}
\end{figure}

The starting point for the development of our theory is that the Lagrangian velocity increment $\bs v = \bs u\left(\bs X(\bs x_0,t_0+\tau),t_0+\tau \right) - \bs u(\bs x_0,t_0)$ can be perceived as an Eulerian two-point-two-time velocity increment between two space-time points which are connected by a Lagrangian trajectory $\bs X(\bs x_0,t_0+\tau)$ starting from $\bs x_0$ at $t_0$. The Lagrangian increment statistics can therefore be related to the Eulerian statistics in an exact manner (see \ref{sisec:generalbridgingrelation} for a detailed derivation):\OK
\begin{equation}\label{eq:exacteulerianlagrangianrelation}
  f^{\mathrm{L}}(\bs v;\tau) = \int\!\mathrm{d}\bs R \, f^{\mathrm{E}}(\bs v; \bs R, \tau) \, P(\bs R | \bs v, \tau) \, .
\end{equation}
Here, $f^{\mathrm{L}}(\bs v;\tau)$ is the PDF of the Lagrangian velocity increment, $f^{\mathrm{E}}(\bs v; \bs R, \tau)$ is the Eulerian two-point-two-time increment PDF, and $P(\bs R | \bs v, \tau)$ is the conditional Lagrangian dispersion PDF of the particle displacement $\bs R = \bs X(\bs x_0,t_0+\tau) - \bs x_0$, which describes how particles spread over time lag $\tau$ depending on the velocity increment $\bs v$. To establish a bridging relation between \emph{temporal} Lagrangian and \emph{instantaneous spatial} Eulerian increment statistics, we need to develop an understanding of the Eulerian temporal decorrelation (contained in the Eulerian two-point-two-time increment PDF) as well as of the Lagrangian single-particle dispersion. 

For long times, Lagrangian tracer particles have typically travelled far from their point of origin.
As the particles are advected with the local instantaneous velocity, all correlations with the velocity increment taken over the travelled distance have decayed, an assumption consistent with Corrsin's hypothesis \cite{corrsin59agp}. As a consequence $P(\bs R | \bs v, \tau)$ becomes independent of $\bs v$ and simplifies to the unconditional standard Lagrangian dispersion PDF. The Lagrangian dispersion PDF is close to Gaussian in homogeneous isotropic turbulence, and therefore fully characterized by its mean squared displacement, which grows diffusively for long times, i.e.~$\langle \bs R^2 \rangle \sim \tau$. While the particles disperse, the flow also decorrelates at their point of origin. The Eulerian temporal decorrelation is primarily governed by the random advection of velocity fluctuations by larger-scale eddies past the fixed Eulerian position, so-called random sweeping \cite{kraichnan64pof,tennekes75jfm}. Based on this hypothesis the Eulerian two-point-two-time increment PDF is obtained by blurring out the instantaneous Eulerian increment PDF with a Gaussian ``random sweeping decorrelation PDF'' (see \ref{sisec:longtimebridgingrelation} for a detailed derivation), whose variance also grows diffusively for long times. The long-time behavior therefore can be captured by combining the two classical ideas of Corrsin's statistical independence hypothesis for the Lagrangian particle dispersion and the Tennekes-Kraichnan random sweeping hypothesis for the Eulerian decorrelation in one effective PDF; we obtain \OK
  $f^{\mathrm{L}}(\bs v;\tau) \approx \int\!\mathrm{d}\bs R \, f^{\mathrm{E}}(\bs v; \bs R) \, P(\bs R ; \tau) \, ,$
where $P(\bs R ; \tau)$ is close to Gaussian with a variance of $\langle \bs R^2 \rangle \sim \tau$ containing contributions from both Lagrangian single-particle dispersion as well as random sweeping decorrelation. In comparison to \eref{eq:exacteulerianlagrangianrelation} this relation features Eulerian single-time information only and shows that particle dispersion and random sweeping introduce a probabilistic sampling of Eulerian velocity fluctuations over a range of spatial scales.

For short times, Corrsin's approximation as well as the assumption that random sweeping decorrelation and particle dispersion are independent are clearly violated. In fact, the latter two effects are induced both by the same advection velocity and happen therefore coherently. In the following, this will be accounted for by removing spurious sweeping effects in the spirit of Kolmogorov's original works \cite{kolmogorov91prs,monin07book}. Furthermore, statistical correlations between particle dispersion and velocity fluctuations play a crucial role at short times.

To determine the short-time behavior, we consider Lagrangian trajectories which have a common initial velocity, visualized in figure \ref{fig:effectivelagrangiandispersion}. Although the particles start with the same initial velocity, the individual trajectories deviate very soon. This virtual blob rapidly deforms and spreads across the volume due to turbulent mixing. The center of mass of this blob of particles, however, drifts in very good approximation ballistically for short times. To see that, we introduce the mean Lagrangian displacement $\overline {\bs X}(\bs x_0,t)-\bs x_0 = \langle {\bs X}(\bs x_0,t) - \bs x_0 | \bs u_0 \rangle$ where the average is performed over all Lagrangian particles with fixed initial velocity $\bs u_0$. For short times, the mean Lagrangian displacement can be expanded in a Taylor series: \OK
\begin{equation}
  \overline {\bs X}(\bs x_0,t) - \bs x_0 =  \bs u_0 t + \frac{1}{2} \langle \bs a_0 | \bs u_0 \rangle t^2 + \mathrm{h.o.t.}
\end{equation}
Here, $\langle \bs a_0 | \bs u_0 \rangle$ denotes the average acceleration at the initial Lagrangian point conditional on the initial velocity.
Because the average is taken over many different flow configurations with fixed initial velocity, we obtain $\langle \bs a_0 | \bs u_0 \rangle \approx \bs 0$ due to a weak directional correlation between the large-scale velocity and the small-scale acceleration field. The same argument applies to the higher-order terms. Therefore, the short-time behavior is given by a ballistic flight. In fact, this Taylor expansion is expected to hold for much longer than the Taylor expansion of an individual Lagrangian trajectory --- there higher-order terms cannot be neglected even on the order of the Kolmogorov time scale, which characterizes the small time scales of the flow.

As the mean Lagrangian trajectory $\overline {\bs X}(\bs x_0,t)$ moves ballistically, it approximately maintains its initial velocity.
As a result the Lagrangian velocity increment at small time lags can be expressed as $\bs v \approx \bs u\left(\bs X(\bs x_0,t_0+\tau),t_0+\tau \right) - \bs u(\overline {\bs X}(\bs x_0, t_0+\tau),t_0+\tau)$ such that velocity increment is now taken at the same time $t_0 + \tau$ over the ``effective Lagrangian displacement'' $\bs R = \bs X(\bs x_0,t_0+\tau)-\overline {\bs X}(\bs x_0,t_0+\tau)$. The instantaneous Eulerian field is therefore sampled on a scale smaller than na\"ively expected based on single-particle dispersion. As a result we obtain a bridging relation very similar to \eref{eq:exacteulerianlagrangianrelation} (see \ref{sisec:shorttimebridgingrelation} for a detailed derivation): \OK
\begin{equation}\label{eq:shorttimebridging}
 f^{\mathrm{L}}(\bs v;\tau) \approx \int\!\mathrm{d}\bs R \, f^{\mathrm{E}}(\bs v; \bs R) \, P(\bs R | \bs v, \tau) \, .
\end{equation}
This relation again features only instantaneous Eulerian information, now mixed by an effective Lagrangian dispersion process which removes sweeping effects.

Comparing the two bridging relations for short and long times, we arrive at the remarkable conclusion that they can be unified by identifying $P(\bs R | \bs v, \tau)$ as the ``effective Lagrangian dispersion PDF'' that combines the dispersion of tracer particles for short times and the combined effects of particle dispersion and random sweeping for long times. We have already established its shape for large time lags as Gaussian with a diffusively increasing variance. For short times, the effective Lagrangian dispersion PDF in comparison has to be sharply localized. Here, the effective Lagrangian displacement is in leading order given by the initial acceleration, $\bs X(\bs x_0,t_0+\tau)-\overline {\bs X}(\bs x_0,t_0+\tau)  \approx \bs a_0 \tau^2/2$. For such small time lags, the acceleration can be expressed in terms of the velocity increment, $\bs a_0 \approx \bs v/\tau$. This exact result, also discussed in \cite{yakhot08arx}, is reminiscent of the phenomenological bridging relation used in the multifractal framework \cite{boffetta02pre,biferale04prl,borgas93ptr,chevillard12crp}. The two limiting cases (short and long time behavior) can be unified into a model for the effective Lagrangian dispersion PDF taking the form:\OK
\begin{equation}\label{eq:ELDPDFmodel}
  P(\bs R | \bs v, \tau ) = \frac{1}{(2\pi)^{3/2}\sigma(\tau)^3} \exp\left[-\frac{\left(\bs R-\mu(\tau) \bs v \right)^2}{2\sigma(\tau)^2} \right] \, ,
\end{equation}
where for short times $\sigma(\tau)$ approaches zero such that the Gaussian asymptotically converges to a delta function while $\mu(\tau)$ behaves like $\tau/2$. For long times, however, $\mu(\tau)$ goes to zero as the correlations with the velocity increment decay, whereas $\sigma(\tau)$ reaches a diffusive regime.

\begin{figure}[t]
\centering
\includegraphics{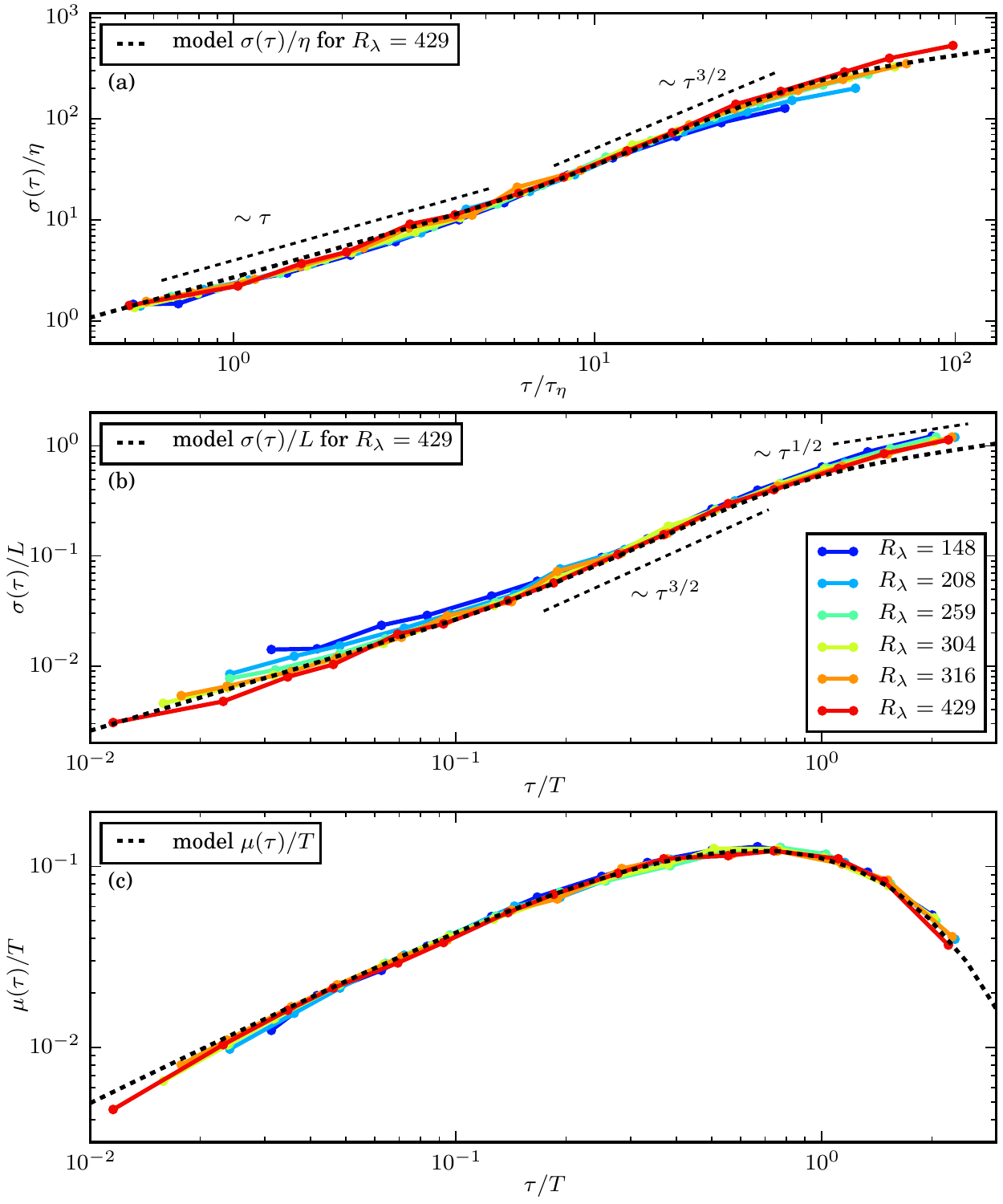}
\caption{Width $\sigma(\tau)$ and shift $\mu(\tau)$ of the angle-integrated effective Lagrangian dispersion PDF \eref{eq:ELDPDFmodelaveraged} obtained from the simulation data.
(a) and (b) show $\sigma(\tau)$ as a function of $\tau/\tau_{\eta}$ and $\tau/T$, respectively.
The collapse at small and large scales, respectively, demonstrates that $\sigma(\tau)$ contains both dissipative and integral-scale contributions.
The dotted line shows the empirical formula \eref{eq:fittedsigma} for $\sigma(\tau)$ for the highest Reynolds number. (c): The shift $\mu(\tau)$ is for small times in very good agreement with the analytical prediction $\tau/2$, and collapse in integral scale units is demonstrated. The dotted line shows the empirical formula \eref{eq:fittedmu} for $\mu(\tau)$.}
\label{fig:sigmamu}
\end{figure}

\section{Results from direct numerical simulations}

To test our theory, we ran an extensive suite of direct numerical simulations (DNS) of homogeneous isotropic turbulence covering a Reynolds number range from $R_{\lambda} \approx 150$ to $R_{\lambda} \approx 430$.
The data was generated with a standard pseudo-spectral code solving the Navier-Stokes equation in the vorticity formulation.
DNS setup and algorithms are described in more detail in \ref{sisec:DNSdescription}.

To further simplify our theoretical predictions we make use of isotropy. As one implication, the full statistical information of the Lagrangian velocity increment vector is contained in the velocity increment magnitude, and the corresponding bridging relation is readily derived (see \ref{sisec:isotropicbridging} for details): \OK
\begin{equation}\label{eq:magnitudebridging}
  f^{\mathrm{L}}(v;\tau) = N(\tau) \int\! \mathrm{d} R \, f^{\mathrm{E}}(v ; R) \, P(R | v, \tau) \, .
\end{equation}
For $P(R | v, \tau)$ we take the angle-integrated version of \eref{eq:ELDPDFmodel}, which takes the form \OK
\begin{equation}\label{eq:ELDPDFmodelaveraged}
 P(R | v,\tau) = \sqrt\frac{2}{\pi} \frac{R^2}{\sigma(\tau)^3}\exp\left[ -\frac{R^2+\mu(\tau)^2v^2}{2\sigma(\tau)^2} \right] \, \mathrm{sinhc}\left[ \frac{\mu(\tau) v R}{\sigma(\tau)^2} \right] \, ,
\end{equation}
where $\mathrm{sinhc}(x) = \mathrm{sinh}(x)/x$. The angle-integrated effective Lagrangian dispersion PDF simplifies to a Maxwellian for $\mu(\tau)=0$, but otherwise displays a non-trivial dependence on the velocity increment. Because the proposed model PDF mixes different Eulerian scales for different velocity increments, we also have to include the factor $N(\tau)$ to keep the model Lagrangian PDFs normalized. For the data sets under consideration $N(\tau)$ is of order unity, as expected. To compare our model to the DNS data we combine \eref{eq:ELDPDFmodelaveraged} with \eref{eq:magnitudebridging} and determine $\mu(\tau)$ and $\sigma(\tau)$ by a least square fitting of the structure functions (moments of the velocity increment PDF) up to order eight. The detailed fitting procedure is explained in \ref{sisec:fitting}. The results are shown in figure \ref{fig:sigmamu}. The width of the effective Lagrangian dispersion PDF is controlled by $\sigma(\tau)$, which is shown in panels (a) and (b), non-dimensionalized by Kolmogorov units and integral scale units, respectively.
Remarkably, non-dimensionalization with the Kolmogorov units leads to the collapse of the curves obtained from various Reynolds number data sets at small scales, whereas non-dimensionalization with integral scales leads to a collapse at large scales. Closer inspection of the results suggests that $\sigma(\tau)$ first increases linearly and then continues to grow approximately like $\tau^{3/2}$. The effective Lagrangian dispersion takes out sweeping effects and describes the two-particle separation between real and the virtual mean Lagrangian trajectories. It is therefore tempting to assume that the $\tau^{3/2}$-growth is due to a Richardson dispersion-like behavior. On theoretical grounds, $\sigma(\tau)$ is expected to reach the long-time limit $\sim u T^{1/2}\tau^{1/2}$, which is in reasonable agreement with $\sigma(\tau)$ obtained from the DNS data. Deviations at the largest time lags are likely rooted in the fact that the fitting procedure becomes less sensitive to changes in $\sigma(\tau)$ as the increment PDFs reach their asymptotic close-to-Gaussian shape.
The cross-over between the regimes depends on the Reynolds number. In fact, we find that this behavior can be captured for all Reynolds numbers in the empirical formula \OK
\begin{equation}\label{eq:fittedsigma}
\sigma(\tau) = 1.95\, u \, \tau \,R_{\lambda}^{-1/3}\left[ 1 + \left(\frac{\tau}{0.853 \, R_{\lambda}^{1/3} \tau_{\eta}}\right)^{4} \right]^{1/8}\left[ 1 + \left(\frac{\tau}{0.895 T}\right)^{4} \right]^{-1/4}  .
\end{equation}
Interestingly, our investigations identify the Eulerian integral time scale as the one controlling the transition at large scales. This could be rooted in the fact that the long-time limit is dominated by random sweeping decorrelation along with the expectation that the Eulerian and Lagrangian integral time scale are roughly of the same order \cite{corrsin63jas}.

The shift of the effective Lagrangian dispersion PDF is controlled by $\mu(\tau)$, which is shown in panel (c) of figure \ref{fig:sigmamu}.
For short times the estimate obtained from the data is in good agreement with our analytical prediction of $\mu(\tau)\approx\tau/2$, whereas $\mu(\tau)$ drops on the order of the integral time scale. When scaled by the integral time scale, the fits for various Reynolds numbers collapse.
For the Reynolds numbers under consideration, we find that \OK
\begin{equation}\label{eq:fittedmu}
  \mu(\tau) = \frac{\tau}{2}\exp\left[ -1.50 \, \frac{\tau}{T} \right]
\end{equation}
captures the numerical observations well. It has to be mentioned though that the exponential decay is delicate to estimate from the data: as $\mu$ becomes less significant, its accurate estimation becomes very difficult.

\begin{figure}[t]
\centering
\includegraphics[width=\linewidth]{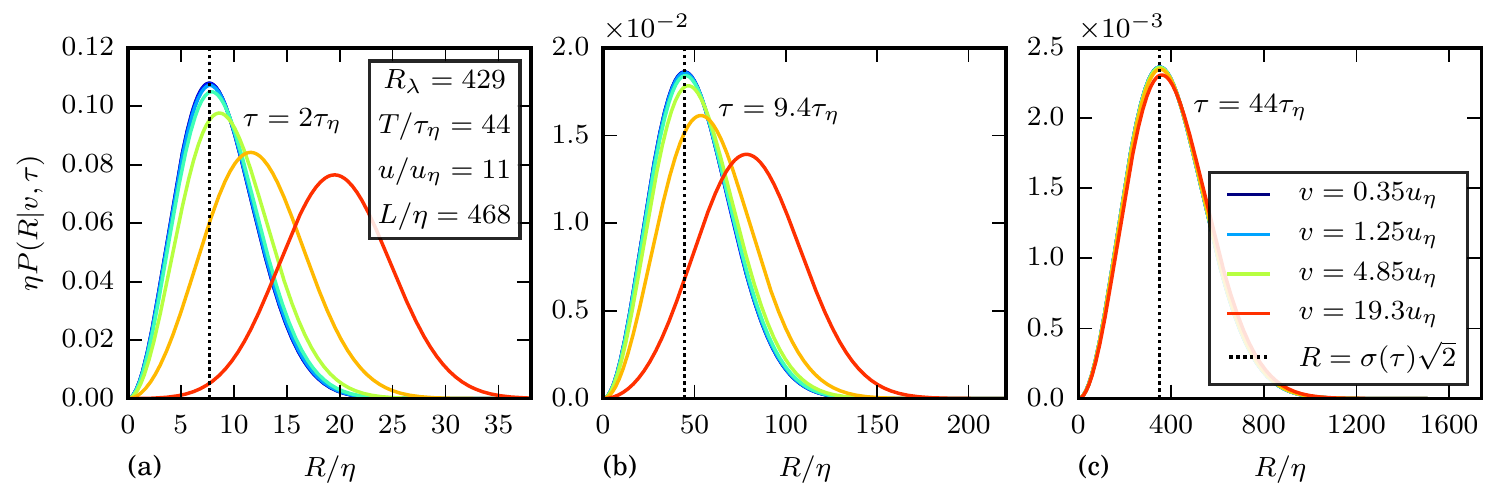}
\caption{Angle-integrated effective Lagrangian dispersion PDF \eref{eq:ELDPDFmodelaveraged} for various time lags with $\sigma(\tau)$ and $\mu(\tau)$ obtained from DNS data. (a)-(b): The variation of the PDF with the velocity increment at short and intermediate times demonstrates that the Eulerian scales are mixed depending on the magnitude of velocity fluctuations. (c): At long times, comparable to the integral time scale, the PDFs collapse to a simple Maxwellian independent of $v$.}
\label{fig:transitionpdfs}
\end{figure}

Figure \ref{fig:transitionpdfs} shows the effective Lagrangian dispersion PDF \eref{eq:ELDPDFmodelaveraged} for various time lags $\tau$ with $\sigma(\tau)$ and $\mu(\tau)$ given by \eref{eq:fittedsigma} and \eref{eq:fittedmu}, respectively. For small to moderate time lags, the PDF displays a pronounced dependence on the velocity increment, as shown in panels (a) and (b): for larger velocity increments, relatively larger Eulerian scales with markedly larger variances are mixed into the Lagrangian statistics. This provides a mechanism to generate heavier tails of the Lagrangian PDFs compared to the Eulerian. This effect weakens with increasing time lag. On the order of the integral time scale, Corrsin's approximation begins to hold and the effective Lagrangian dispersion PDF becomes independent of the velocity increment; as a result it simplifies to a standard Maxwellian, as demonstrated in panel (c). In essence, the effect of the bridging relation is two-fold: Eulerian statistics of various scales are mixed to obtain Lagrangian statistics. Additionally, the scale-mixing depends on the magnitude of the velocity fluctuations. This allows for a different statistical behavior of higher-order statistics, dominated by the tails of the PDF, compared to lower orders, dominated by its core.

\begin{figure}
\centering
\includegraphics[width=\linewidth]{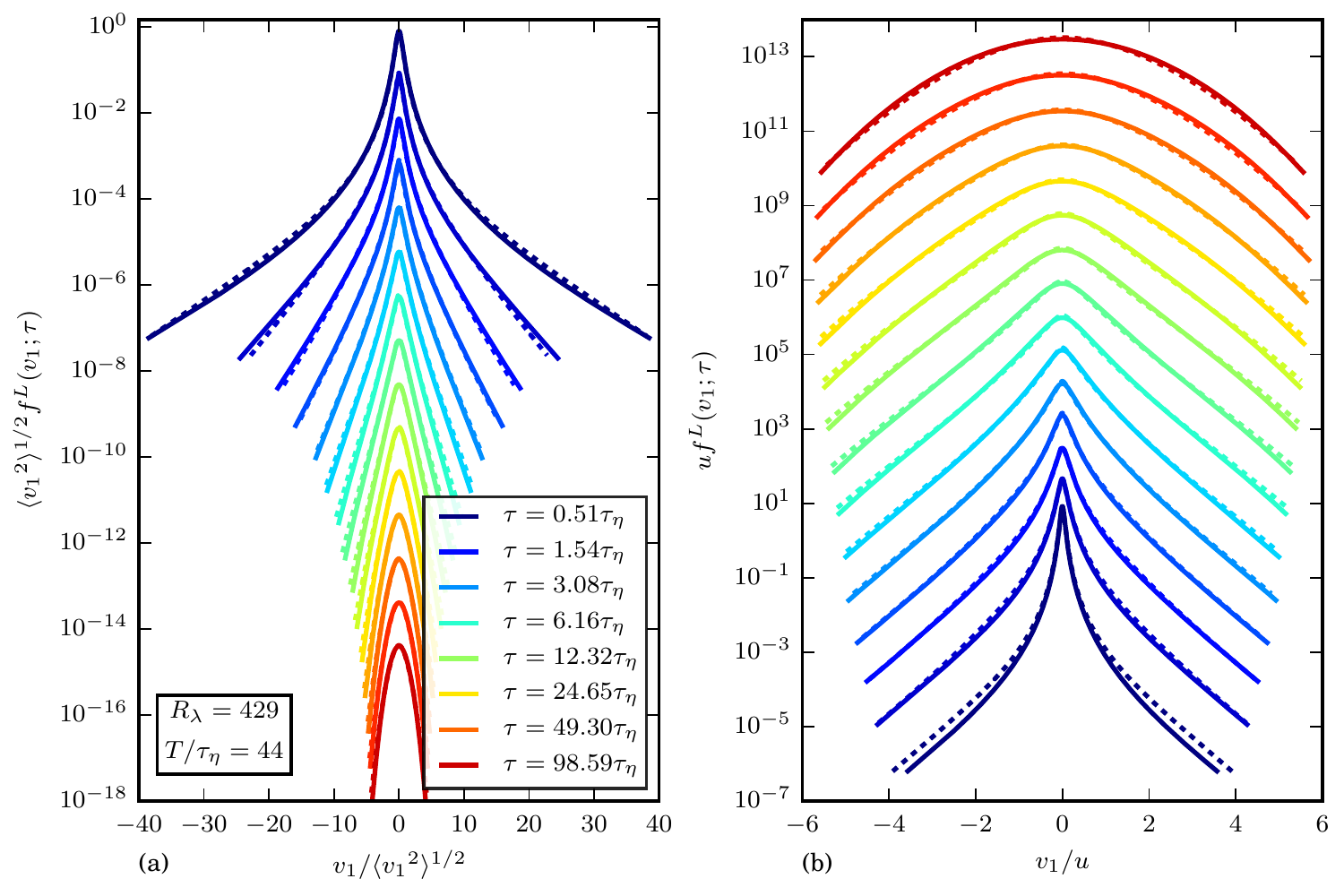}
\caption{Lagrangian velocity increment PDFs from simulations (continuous lines) along with PDFs obtained from the bridging relation based on Eulerian increment PDFs (dotted lines). Panel (a) shows the PDFs scaled to unit standard deviation, panel (b) shows the unstandardized PDFs. 
Excellent agreement is found from about one Kolmogorov time scale on to the integral time scale and beyond.}
\label{fig:incrementpdfs}
\end{figure}

To test the fidelity of the bridging relation, we obtain Eulerian velocity increment magnitude PDFs from the DNS and compute Lagrangian increment magnitude PDFs using the angle-integrated version of the effective Lagrangian dispersion PDF along with the formulas for $\sigma(\tau)$ and $\mu(\tau)$. As it is more common to consider PDFs of the velocity increment components rather than the PDFs of magnitudes, we compute the standard component velocity increment PDFs from \OK
\begin{equation}
  f^L(v_1;\tau) = \int_{|v_1|}^{\infty}\!\!\!\!\!\mathrm{d}v \, \frac{f^L(v;\tau)}{2v} \, .
\end{equation}
Figure \ref{fig:incrementpdfs} shows a comparison of the increment PDFs directly estimated from the numerical data (solid lines) with the results from the bridging relation (dashed lines) for the highest Reynolds number case. Excellent agreement from the dissipative scales starting from about one Kolmogorov time scale all the way up to and beyond the integral time scale is found. The visible deviations below one Kolmogorov time scale could be either systematic or caused by the limited scale-resolution of the Eulerian increment PDFs, which are used to evaluate the bridging relation. A detailed investigation of the dissipative-range physics of the bridging relation, including an extension to acceleration statistics, is the subject of future work.

\section{Conclusions and Outlook}

In summary, we have introduced a novel bridging relation that relates Lagrangian temporal velocity fluctuations to instantaneous Eulerian spatial velocity fluctuations. We showed that Lagrangian increment statistics can be perceived as a scale-mixing of Eulerian increment statistics by an effective Lagrangian dispersion. Based on the physics of single-particle transport as well as random sweeping effects, we derived the effective Lagrangian dispersion PDF and determined its asymptotic behavior. The predictions for the Lagrangian velocity increment PDFs were found to be in excellent agreement with results from high-resolution direct numerical simulations of fully developed turbulence.

Supported by the extensive numerical investigations, we conclude that for the Reynolds number range under consideration Lagrangian intermittency can be captured with the effective Lagrangian dispersion PDF, whose shape is independent of Reynolds number. The Reynolds number dependence enters through the Eulerian statistics and through the width $\sigma(\tau)$. The fact that the effective Lagrangian dispersion mixes Eulerian statistics across a wide range of scales indicates that Lagrangian velocity increment statistics typically contain both dissipative as well as integral-scale contributions (see also the related discussion in \cite{kamps09pre}). This may provide an explanation of a reduced scaling range of Lagrangian structure functions compared to their Eulerian counterparts; for instance, it was argued that even for the second-order (i.e. the lowest non-trivial order) Lagrangian structure function a clear scaling range is only expected starting at $R_{\lambda}\approx 5 \cdot 10^3$ \cite{lanotte13jot} or even beyond  $R_{\lambda}\approx 3 \cdot 10^4$ \cite{sawford11pof}, where Eulerian scaling is already well established.

The presented approach is very general and therefore opens avenues to a new generation of models for transport in complex particle-laden flows by allowing an assessment of particle statistics from the knowledge of the flow field statistics only. For instance, an exciting future direction would be the extension of the theory to inertial and finite-sized particles. Here, the effect of preferential concentration will imply a biased sampling of the flow leading to modifications of the effective Lagrangian dispersion PDF. It also appears worthwhile to pursue applications to atmospheric flows and turbulent convection by including characteristic large-scale features of these flows.  Finally, it will be most interesting to go beyond hydrodynamic flows and investigate the bridging induced by the effective Lagrangian dispersion PDF, for example, in magnetohydrodynamic flows, which display a fine-scale structure markedly different from hydrodynamic turbulence \cite{busse07pop,homann09njp}.

\section*{Acknowledgments}
We gratefully acknowledge stimulating discussions with O.~Kamps, J.~Friedrich and R.~Grauer as well as helpful comments by A.~Celani, M.~Cencini and A.~Scagliarini. We thank P.~Johnson, J.~Lawson and M.~Sinhuber for comments on the manuscript as well as insightful discussions. Computations were performed on the clusters of the Max Planck Computing and Data Facility. This work was supported by the Max Planck Society.

\appendix
\section{}

\subsection{Theoretical background: the bridging relation}
\label{sisec:generalbridgingrelation}

To establish the probabilistic laws connecting Eulerian and Lagrangian statistics, we consider a velocity field $\bs u(\bs x,t)$ of a homogeneous isotropic turbulent flow governed by the incompressible Navier-Stokes equation in the statistically stationary state. A very small (compared to the smallest scales of the flow) neutrally buoyant Lagrangian tracer particle starting from point $\bs x_0$ at time $t_0$ is advected by the velocity field according to \OK
\begin{equation}
  \frac{\mathrm{d}\phantom{t}}{\mathrm{d} t}\bs X(\bs x_0,t) = \bs u\left(\bs X(\bs x_0,t),t \right)
\end{equation}
with the initial condition $\bs X(\bs x_0,t_0) = \bs x_0$.
As the particle probes the flow over the time lag $\tau$, it encounters velocity fluctuations which can be characterized by the Lagrangian velocity increment $\bs u\left(\bs X(\bs x_0,t_0+\tau),t_0+\tau \right) - \bs u(\bs x_0,t_0)$.
The statistics of the velocity fluctuations can be conveniently captured in terms of the probability density function (PDF) \OK
\begin{equation}\label{eq:lagincpdf_SI}
  f^{\mathrm{L}}(\bs v;\tau) = \left\langle \delta\left[ \bs u\left(\bs X(\bs x_0,t_0+\tau),t_0+\tau \right) - \bs u(\bs x_0,t_0) - \bs v \right] \right\rangle \, .
\end{equation}
Here, $\langle \dots \rangle$ denotes an ensemble average over all Lagrangian particles and all velocity fields that can also be replaced with a spatio-temporal average for practical purposes.
The PDF $f^{\mathrm{L}}(\bs v;\tau)$ is a normalized density with respect to the sample-space variable $\bs v$ with a functional dependence on the time lag $\tau$. Note that it does not explicitly depend on the initial time $t_0$ due to statistical stationarity. Eq.~\eref{eq:lagincpdf_SI} essentially contains statistical information of a Lagrangian tracer particle sampling the Eulerian, fixed lab-frame velocity fluctuations at two  points in space and time. In fact, this statement can be made rigorous by considering \OK
\begin{eqnarray}
\fl  f^{\mathrm{L}}(\bs v;\tau) = \left\langle \delta\left[ \bs u\left(\bs X(\bs x_0,t_0+\tau),t_0+\tau \right) - \bs u(\bs x_0,t_0) - \bs v \right] \right\rangle \label{eq:eulerian_lagrangian_bridging1} \\
  = \int\!\mathrm{d}\bs y \left\langle \delta\left[ \bs u\left(\bs y,t_0+\tau \right) - \bs u(\bs x_0,t_0) - \bs v \right] \, \delta[\bs X(\bs x_0,t_0+\tau)-\bs y] \right\rangle \label{eq:eulerian_lagrangian_bridging2} \\
  = \int\!\mathrm{d}\bs y \left\langle \delta\left[ \bs u\left(\bs y,t_0+\tau \right) - \bs u(\bs x_0,t_0) - \bs v \right] \right\rangle \left\langle \delta[\bs X(\bs x_0,t_0+\tau)-\bs y] \big | \bs v \right\rangle  \label{eq:eulerian_lagrangian_bridging3} \\
  = \int\!\mathrm{d}\bs y \, f^{\mathrm{E}}(\bs v ; \bs y - \bs x_0, \tau) \left\langle \delta[\bs X(\bs x_0,t_0+\tau)-\bs x_0-\bs y+\bs x_0] \big | \bs v \right\rangle  \label{eq:eulerian_lagrangian_bridging4} \\
  = \int\!\mathrm{d}\bs R \, f^{\mathrm{E}}(\bs v ; \bs R, \tau) \left\langle \delta[\bs X(\bs x_0,t_0+\tau)-\bs x_0-\bs R] \big | \bs v \right\rangle  \label{eq:eulerian_lagrangian_bridging5} \\
  = \int\!\mathrm{d}\bs R \, f^{\mathrm{E}}(\bs v ; \bs R, \tau) \, P(\bs R | \bs v, \tau) \, .  \label{eq:eulerian_lagrangian_bridging6}
\end{eqnarray}
From \eref{eq:eulerian_lagrangian_bridging1} to \eref{eq:eulerian_lagrangian_bridging2} the Eulerian point $\bs y$ has been introduced. From \eref{eq:eulerian_lagrangian_bridging2} to \eref{eq:eulerian_lagrangian_bridging3} the average has been split into the Eulerian increment PDF and the PDF of the Lagrangian particle position conditional on the velocity increment. In \eref{eq:eulerian_lagrangian_bridging4} homogeneity and stationarity have been exploited for the Eulerian increment PDF. As a result, it only depends on the time lag $\tau$ and the spatial difference $\bs y - \bs x_0$. Also the starting point $\bs x_0$ has been added and subtracted in the argument of the delta function of the Lagrangian particle position. From \eref{eq:eulerian_lagrangian_bridging4} to \eref{eq:eulerian_lagrangian_bridging5} a change of variables was applied, which finally allows us to identify the conditional Lagrangian dispersion PDF in \eref{eq:eulerian_lagrangian_bridging6}. This concludes the derivation of \eref{eq:exacteulerianlagrangianrelation} from the main text.

The above bridging relation still contains Eulerian \emph{two-time} information. The goal in the following is to establish a relation between $f^{\mathrm{L}}(\bs v;\tau)$ and the Eulerian \emph{single-time} velocity increment PDF \OK
\begin{equation}
  f^{\mathrm{E}}(\bs v;\bs r) = \left\langle \delta\left[ \bs u\left(\bs x + \bs r,t\right) - \bs u(\bs x,t) - \bs v \right] \right\rangle \,.
\end{equation}
Due to homogeneity, this PDF does not explicitly depend on the spatial coordinate $\bs x$, and, as before, it does not depend explicitly on the time $t$ due to statistical stationarity. The main challenge here is to relate the Eulerian two-time statistics to the single-time statistics. As outlined in the main text, the temporal effects can be absorbed into the Lagrangian dispersion PDF by introducing an effective dispersion. The physical arguments therefor are given in the main text, here we add some technical details.

\subsection{Bridging relation for long times}
\label{sisec:longtimebridgingrelation}

We show here that the long-time behavior is captured by the proposed bridging relation. For very long times we obtain \OK
{\footnotesize
\begin{eqnarray}
\fl f^{\mathrm{L}}(\bs v;\tau) = \int\!\mathrm{d}\bs R \left\langle \delta\left[ \bs u\left(\bs x_0 + \bs R,t_0+\tau \right) - \bs u(\bs x_0,t_0) - \bs v \right] \, \delta[\bs X(\bs x_0,t_0+\tau)-\bs x_0-\bs R]  \right\rangle \label{eq:longbridging1} \\
\fl  \approx \int\!\mathrm{d}\bs R \left\langle \delta\left[ \bs u\left(\bs x_0 + \bs R,t_0+\tau \right) - \bs u(\bs x_0,t_0) - \bs v \right] \right\rangle \left\langle \delta[\bs X(\bs x_0,t_0+\tau)-\bs x_0-\bs R]  \right\rangle \label{eq:longbridging2} \\
\fl  = \int\!\mathrm{d}\bs R \, f^{\mathrm{E}}(\bs v ; \bs R,\tau) \, f^{\mathrm{LD}}(\bs R ; \tau ) \, .
\end{eqnarray}}
Here, we have introduced the standard Lagrangian dispersion PDF $f^{\mathrm{LD}}(\bs R ; \tau ) = \left\langle \delta[\bs X(\bs x_0,t_0+\tau)-\bs x_0-\bs R]  \right\rangle$ and the Eulerian two-point-two-time increment PDF $f^{\mathrm{E}}(\bs v ; \bs R,\tau)$ and assumed that the two PDFs become statistically independent for long times, which is essentially Corrsin's hypothesis and well justified by the physics of the problem.

The Eulerian two-point-two-time increment PDF can be related to the single-time increment PDF by applying the random sweeping hypothesis \cite{kraichnan64pof,tennekes75jfm}, i.e.~we assume that the temporal decorrelation is dominated by large-scale random advection effects. That means, velocities are carried past a fixed point in space with a random sweeping velocity over a distance $\bs R_{\mathrm{sw}}(\tau) = \int_0^{\tau}\mathrm{d}s \, \bs u_{\mathrm{sw}}(s)$, where $\bs u_{\mathrm{sw}}$ is the large-scale random-sweeping velocity at the fixed Eulerian position. This allows us to map the velocity back in time according to \OK
\begin{equation}
  \bs u\left( \bs x, \tau \right) \approx \bs u\left( \bs x-\bs R_{\mathrm{sw}}(\tau) , 0 \right) \, .
\end{equation}
For simplicity we assume that $\bs u_\mathrm{sw}$ is an isotropic Gaussian process with zero mean and correlation function $\langle \bs u_{\mathrm{sw}}(t) \cdot \bs u_{\mathrm{sw}}(t') \rangle$/3. As a consequence, the random sweeping displacement is also a Gaussian process with \OK
\begin{equation}
  \left\langle \bs R_{\mathrm{sw}}^2  \right\rangle(\tau) = \int_0^{\tau} \!\!\mathrm{d}s \int_0^{\tau} \!\! \mathrm{d}s' \, \langle \bs u_{\mathrm{sw}}(s) \cdot \bs u_{\mathrm{sw}}(s') \rangle \, .
\end{equation}
 As most of the kinetic energy is contained in the large scales of the flow, we have $\langle \bs u_{\mathrm{sw}}^2 \rangle/3 \approx \langle \bs u^2 \rangle$/3 for the single-time variance. For short times, the mean squared displacement grows ballistically, $\left\langle \bs R_{\mathrm{sw}}^2  \right\rangle(\tau) \approx \langle \bs u^2 \rangle \tau^2$. For long times, significantly larger than the Eulerian integral time scale $T$, the displacement becomes diffusive, i.e. $\left\langle \bs R_{\mathrm{sw}}^2  \right\rangle(\tau) \approx 2\langle \bs u^2 \rangle T \tau$. Taking into account these considerations,  we can relate the two-time PDF to the single-time PDF via \OK
\begin{eqnarray}
 f^{\mathrm{E}}(\bs v; \bs r, \tau) = \left\langle \delta\left[ \bs u(\bs x + \bs r, t+\tau) - \bs u(\bs x,t) - \bs v \right]\right\rangle \\
 \approx \left\langle \delta\left[ \bs u(\bs x + \bs r - \bs R_{\mathrm{sw}}(\tau), t) - \bs u(\bs x,t) - \bs v \right]\right\rangle \\
    \approx \int \!\mathrm{d}\bs y \, \left\langle \delta\left[ \bs u(\bs x + \bs r - \bs y, t) - \bs u(\bs x,t) - \bs v \right]\right\rangle \left\langle \delta\left[ \bs R_{\mathrm{sw}}(\tau) - \bs y \right] \right\rangle \\
    = \int \!\mathrm{d}\bs y \, f^{\mathrm{E}}(\bs v; \bs r - \bs y) \, f^{\mathrm{sw}}(\bs y;\tau) \, ,
\end{eqnarray}
where $f^{\mathrm{sw}}$ is the Gaussian random sweeping PDF with variance $\left\langle \bs R_{\mathrm{sw}}^2  \right\rangle(\tau)/3$. In the derivation we have made the assumption of statistical independence of the sweeping velocity and the velocity increment. Combining this result with the long-time bridging relation, we obtain \OK
\begin{eqnarray}
  f^{\mathrm{L}}(\bs v;\tau) &\approx \int\!\mathrm{d}\bs R \,\mathrm{d}\bs y \, f^{\mathrm{E}}(\bs v ; \bs R- \bs y) \, f^{\mathrm{sw}}(\bs y;\tau) \, f^{\mathrm{LD}}(\bs R ; \tau ) \\
  &= \int\!\mathrm{d} \widetilde{\bs R} \, f^{\mathrm{E}}(\bs v ; \widetilde{\bs R}) \int\!\mathrm{d} \bs R \, f^{\mathrm{sw}}(\bs R-\widetilde{\bs R};\tau) \, f^{\mathrm{LD}}(\bs R ; \tau ) \\
  &= \int\!\mathrm{d}\widetilde{\bs R} \, f^{\mathrm{E}}(\bs v; \widetilde{\bs R}) \, P(\widetilde{\bs R} ; \tau) \, .
\end{eqnarray}
The convolution of the Lagrangian dispersion PDF and the random sweeping PDF yields the effective Lagrangian dispersion PDF, showing that the structure of the bridging relation holds for long times. Both, the random sweeping PDF as well as the standard Lagrangian dispersion PDF are in good approximation Gaussian with diffusively growing variances for large times. Therefore also the effective Lagrangian dispersion PDF is Gaussian with a variance $\sim \tau $ for long times.

\subsection{Bridging relation for short times}
\label{sisec:shorttimebridgingrelation}

To make contact to the Eulerian single-time statistics for short times, the mean Lagrangian displacement $\overline {\bs X}(\bs x_0, t) - \bs x_0 = \langle {\bs X}(\bs x_0, t) | \bs u_0 \rangle - \bs x_0 $ has been introduced in the main text. Because Lagrangian particle transport is dominated by random advection by the large-scale velocity, the velocity is expected to vary only little along the mean Lagrangian trajectory and hence $\bs u(\overline {\bs X}(\bs x_0, t_0+\tau),t_0+\tau) = \bs u(\bs x_0,t_0) + \bs u'$, where the correction $\bs u'$ is negligible for small to moderate times. This allows us to approximately identify the initial velocity of the Lagrangian increment with the velocity along the mean Lagrangian trajectory. As a result the bridging relation becomes \OK
\begin{equation}
\fl  f^{\mathrm{L}}(\bs v;\tau) \approx \left\langle \delta\left[ \bs u\left(\bs X(\bs x_0,t_0+\tau),t_0+\tau \right) - \bs u(\overline {\bs X}(\bs x_0,t_0+\tau),t_0+\tau) - \bs v \right] \right\rangle
\end{equation}
for $\tau$ not too large.
The crucial point is that this expression involves two points at a single time only. The remaining steps to the complete bridging relation are now straightforward: \OK
{\footnotesize
\begin{eqnarray}
\fl  f^{\mathrm{L}}(\bs v;\tau) \approx \left\langle \delta\left[ \bs u\left(\bs X(\bs x_0,t_0+\tau),t_0+\tau \right) - \bs u(\overline {\bs X}(\bs x_0,t_0+\tau),t_0+\tau) - \bs v \right] \right\rangle \label{eq:bridging1} \\
\fl  = \int\!\mathrm{d}\bs y \, \mathrm{d}\bs z \left\langle \delta\left[ \bs u\left(\bs y,t_0+\tau \right) - \bs u(\bs z,t_0+\tau) - \bs v \right] \, \delta[\bs X(\bs x_0,t_0+\tau)-\bs y] \, \delta[\overline {\bs X}(\bs x_0,t_0+\tau)-\bs z] \right\rangle \label{eq:bridging2} \\
\fl  = \int\!\mathrm{d}\bs y \, \mathrm{d}\bs z \left\langle \delta\left[ \bs u\left(\bs y,t_0+\tau \right) - \bs u(\bs z,t_0+\tau) - \bs v \right] \right\rangle \left\langle \delta[\bs X(\bs x_0,t_0+\tau)-\bs y] \, \delta[\overline {\bs X}(\bs x_0,t_0+\tau)-\bs z] \big | \bs v \right\rangle  \label{eq:bridging3} \\
\fl  = \int\!\mathrm{d}\bs y \, \mathrm{d}\bs z \, f^{\mathrm{E}}(\bs v ; \bs y - \bs z) \left\langle \delta[\bs X(\bs x_0,t_0+\tau)-\bs y] \, \delta[\overline {\bs X}(\bs x_0,t_0+\tau)-\bs z] \big | \bs v \right\rangle  \label{eq:bridging4} \\
\fl  = \int\!\mathrm{d}\bs y \, \mathrm{d}\bs z \, \mathrm{d}\bs R \, f^{\mathrm{E}}(\bs v ; \bs y - \bs z) \left\langle \delta[\bs X(\bs x_0,t_0+\tau)-\bs y] \, \delta[\overline {\bs X}(\bs x_0,t_0+\tau)-\bs z] \, \delta[\bs y - \bs z - \bs R] \big | \bs v \right\rangle  \label{eq:bridging5} \\
\fl  = \int\! \mathrm{d}\bs R \, f^{\mathrm{E}}(\bs v ; \bs R) \left\langle \delta[\bs X(\bs x_0,t_0+\tau) - \overline {\bs X}(\bs x_0,t_0+\tau) - \bs R] \big | \bs v \right\rangle  \label{eq:bridging6} \\
\fl  = \int\! \mathrm{d}\bs R \, f^{\mathrm{E}}(\bs v ; \bs R) \, P(\bs R | \bs v, \tau ) \, .  \label{eq:bridging7}
\end{eqnarray}}
From \eref{eq:bridging1} to \eref{eq:bridging2} the two Eulerian points $\bs y$ and $\bs z$ have been introduced. From \eref{eq:bridging2} to \eref{eq:bridging3} the average has been split into the Eulerian increment PDF and the joint PDF of the Lagrangian trajectory and the mean Lagrangian trajectory conditional on the velocity increment.
In \eref{eq:bridging4} homogeneity and stationarity have been exploited for the Eulerian increment PDF. As a result, it does not explicitly depend on time and is only a function of the difference $\bs y - \bs z$.
Next, the identity $\int\! \mathrm{d}\bs R \, \delta[\bs y - \bs z - \bs R] =1$ has been inserted, which allows us to introduce the effective Lagrangian dispersion PDF in \eref{eq:bridging6} and \eref{eq:bridging7}.
The main limitation of the short-time approximation is the assumption that the correction $\bs u'$ is negligible.
Including a finite correction $\bs u'$ into the previous derivation necessitates additional assumptions about the statistics of the correction.
However, irrespective of the details of the statistics, the main effect will be an additional blurring of Eulerian statistics.
This can essentially be absorbed into the effective Lagrangian dispersion PDF, making it a robust approximation.
This concludes the derivation of the bridging relation \eref{eq:shorttimebridging} from the main text.

\subsection{Simplifications for isotropic turbulence}
\label{sisec:isotropicbridging}
Up to now we have introduced the bridging relation in its most general form for a vectorial velocity increment $\bs v$. For statistically isotropic flows, the statistics of the vector are entirely determined by the statistics of its magnitude, which simplifies the subsequent analytical and numerical treatment. Here, we explicitly derive the short-time bridging relation for the velocity magnitude PDF, in complete analogy to the derivation \eref{eq:bridging1}-\eref{eq:bridging7}:\OK
{\footnotesize
\begin{eqnarray}
\fl  f^{\mathrm{L}}(v;\tau) \approx \left\langle \delta\left[ | \bs u\left(\bs X(\bs x_0,t_0+\tau),t_0+\tau \right) - \bs u(\overline {\bs X}(\bs x_0,t_0+\tau),t_0+\tau)| - v \right] \right\rangle \label{eq:magnitudebridging1} \\
\fl  = \int\!\mathrm{d}\bs y \, \mathrm{d}\bs z \left\langle \delta\left[ |\bs u\left(\bs y,t_0+\tau \right) - \bs u(\bs z,t_0+\tau)| - v \right] \, \delta[\bs X(\bs x_0,t_0+\tau)-\bs y] \, \delta[\overline {\bs X}(\bs x_0,t_0+\tau)-\bs z] \right\rangle \label{eq:magnitudebridging2} \\
\fl  = \int\!\mathrm{d}\bs y \, \mathrm{d}\bs z \left\langle \delta\left[ |\bs u\left(\bs y,t_0+\tau \right) - \bs u(\bs z,t_0+\tau)| - v \right] \right\rangle \left\langle \delta[\bs X(\bs x_0,t_0+\tau)-\bs y] \, \delta[\overline {\bs X}(\bs x_0,t_0+\tau)-\bs z] \big | v \right\rangle  \label{eq:magnitudebridging3} \\
\fl  = \int\!\mathrm{d}\bs y \, \mathrm{d}\bs z \, f^{\mathrm{E}}(v ; |\bs y - \bs z|) \left\langle \delta[\bs X(\bs x_0,t_0+\tau)-\bs y] \, \delta[\overline {\bs X}(\bs x_0,t_0+\tau)-\bs z] \big | v \right\rangle  \label{eq:magnitudebridging4} \\
\fl  = \int\!\mathrm{d}\bs y \, \mathrm{d}\bs z \, \mathrm{d}\bs R \, f^{\mathrm{E}}(v ; |\bs y - \bs z|) \left\langle \delta[\bs X(\bs x_0,t_0+\tau)-\bs y] \, \delta[\overline {\bs X}(\bs x_0,t_0+\tau)-\bs z] \, \delta[\bs y - \bs z - \bs R] \big | v \right\rangle  \label{eq:magnitudebridging5} \\
\fl  = \int\! \mathrm{d}\bs R \, f^{\mathrm{E}}(v ; R) \left\langle \delta[\bs X(\bs x_0,t_0+\tau) - \overline {\bs X}(\bs x_0,t_0+\tau) - \bs R] \big | v \right\rangle  \label{eq:magnitudebridging6} \\
\fl  = N(\tau)\int\! \mathrm{d} R \, f^{\mathrm{E}}(v ; R) \, P(R | v, \tau ) \, .  \label{eq:magnitudebridging7}
\end{eqnarray}}
The main difference is that we have exploited the fact that $f^{\mathrm{E}}(v ; R)$ depends only on the distance magnitude due to isotropy. As a result, we can integrate over the angles from \eref{eq:magnitudebridging6} to \eref{eq:magnitudebridging7}. In \eref{eq:magnitudebridging7} we have thereby introduced the effective Lagrangian dispersion PDF of the magnitude $R$. If a model PDF is introduced for $P(R | v, \tau )$, such as the one discussed in the main text, the additional factor $N(\tau)$ ensures normalization. The corresponding long-time bridging relation can be straightforwardly obtained in the same manner. This concludes the derivation of the bridging relation \eref{eq:magnitudebridging} from the main text.

\subsection{Description of the DNS}
\label{sisec:DNSdescription}

For this study, an extensive suite of direct numerical simulations of statistically stationary fully developed turbulence in a periodic domain was conducted. A standard MPI-parallel pseudo-spectral scheme is used to solve the vorticity formulation of the Navier-Stokes equations in a periodic cubic box of length $2\pi$ (DNS units).
Aliasing errors are controlled by using a high-order Fourier smoothing \cite{hou07jcp}.
Time stepping is performed by means of a memory-saving third-order Runge-Kutta method \cite{shu88jcp}.
To maintain the statistically stationary state, a large-scale band-passed Lundgren forcing \cite{Lundgren03arb} is applied in the wavenumber range $[1.5, 3]$ (DNS units) with an amplitude of $0.5$ (DNS units).

To obtain the Lagrangian data, tracer particles are advected with the flow.
Cubic splines are used to interpolate the velocity fields.
Time stepping for the particles is performed by means of a fourth-order Adams-Bashforth method (see, e.g., \S6.7 in \cite{atkinson89book}).

As preparation, the Eulerian flow field is advanced in time until a statistically stationary state is obtained.
The production runs are launched using this state as an initial condition, and Lagrangian tracers initialized at random locations in the cubic domain (with a uniform distribution).
The main simulation parameters obtained for the production runs are summarized in table \ref{tab:dnsparameters}.

\subsection{Description of the fitting procedure}
\label{sisec:fitting}

\begin{table}
\begin{center}
\begin{tabular}{cccccccc}
$n$ & $R_{\lambda}$ & $u$ & $L$ & $L / \eta$ & $T / \tau_{\eta}$ & $\frac{t_1-t_0}{T}$ & $k_{\max}\eta$\\
\hline
$1024$ &        $148$ &    $0.95$ &    $0.71$ &    $104$ &    $16.8$ &    $12$ &    $2.8$\\
$1024$ &        $208$ &    $0.96$ &    $0.68$ &    $168$ &    $22.9$ &    $14$ &    $1.7$\\
$1536$ &        $259$ &    $0.97$ &    $0.69$ &    $228$ &    $27.9$ &    $6.2$ &    $1.9$\\
$1536$ &        $304$ &    $0.94$ &    $0.68$ &    $297$ &    $33.5$ &    $6.1$ &    $1.4$\\
$2048$ &        $316$ &    $0.97$ &    $0.71$ &    $292$ &    $32.3$ &    $4.5$ &    $2.0$\\
$2048$ &        $429$ &    $1.0$ &    $0.68$ &    $468$ &    $44.4$ &    $4.4$ &    $1.2$\\
\hline
\end{tabular}
\end{center}
\label{tab:dnsparameters}
\caption{
Main DNS parameters.
Direct numerical simulations are run over the time interval $[t_0, t_1]$; field information is computed on a real space grid of $n^3$ points and $10^7$ tracer trajectories are integrated for each flow.
Characteristic parameters are as follows: Taylor-based Reynolds number $R_{\lambda}$, root-mean-squared velocity $u = (2\int \mathrm{d}k E(k) /3)^{1/2}$ (given in DNS units, $E(k)$ is the energy spectrum), integral length scale $L=\frac{\pi}{2u^2} \int \frac{\mathrm{d}k}{k} E(k)$ (given in DNS units), ratios of integral to Kolmogorov scales ($\eta$ and $\tau_\eta$ are computed from the mean kinetic energy dissipation $\varepsilon$ and the kinematic viscosity $\nu$, and the integral time scale is computed as $T=L/u$), and resolution criterion $k_{\max}\eta$, where $k_{\max}$ is the highest wavenumber resolved by our pseudo-spectral code.}
\end{table}

\begin{figure}[t]
\centering
\includegraphics{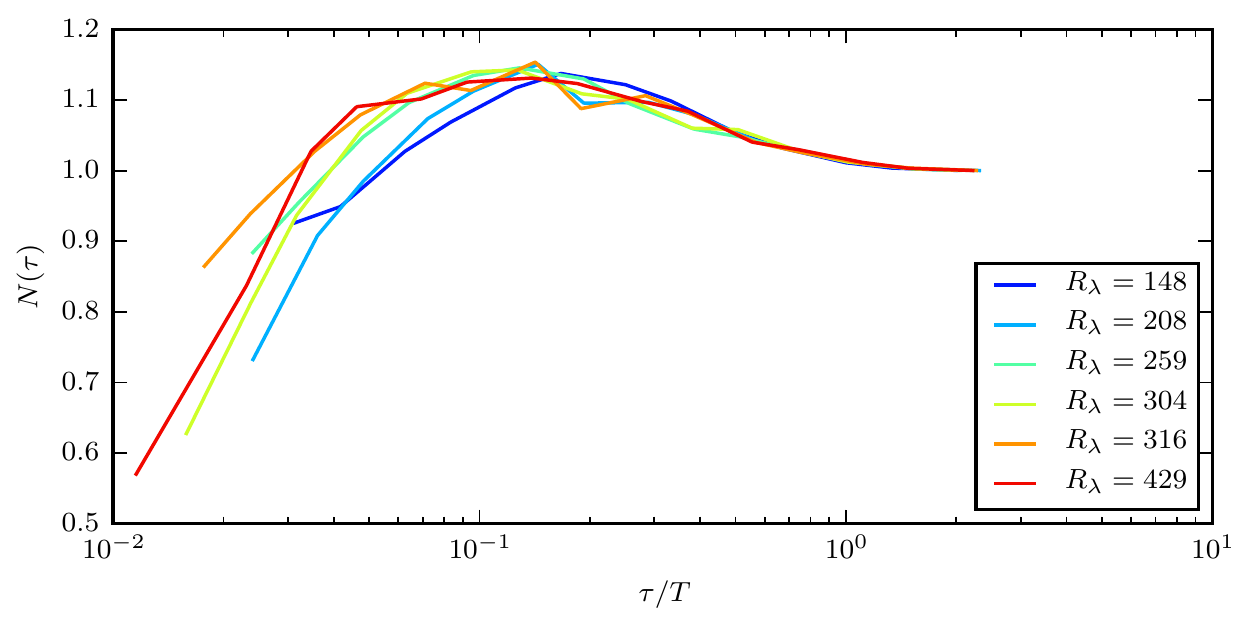}
\caption{
Normalization factor $N(\tau)$ required by \eref{eq:magnitudebridging} for the data at different $R_\lambda$.
$N(\tau)$ reaches unity at the integral time $T$ independently of $R_\lambda$.}
\label{fig:Ntau}
\end{figure}

\begin{figure}[t]
\centering
\includegraphics{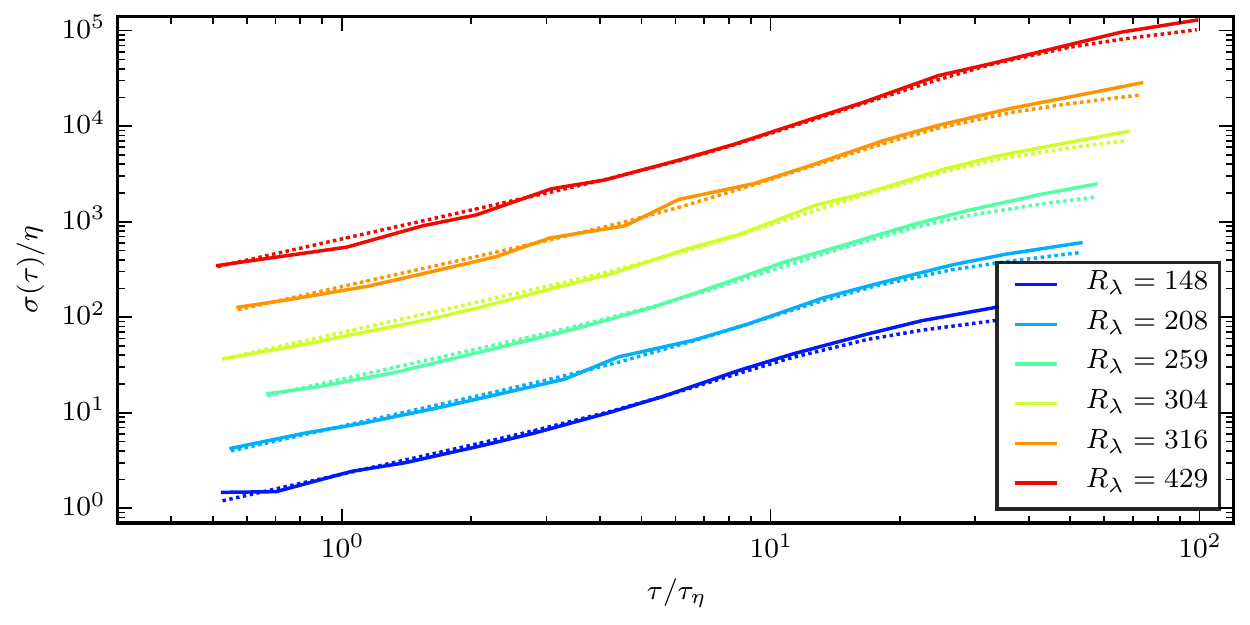}
\caption{
Results of the fitting procedure (continuous lines) compared with the empirical formula for $\sigma(\tau)$ (dotted lines) for various Reynolds numbers. Data from different data sets was multiplied with successive powers of $3$ for clarity.
The data for all the different $R_\lambda$ is matched quite well by the empirical three-power-law formula \eref{eq:fittedsigma} for $\sigma(\tau)$ given in the main text.
}
\label{fig:fitsigma}
\end{figure}

\begin{figure}[h]
\centering
\includegraphics{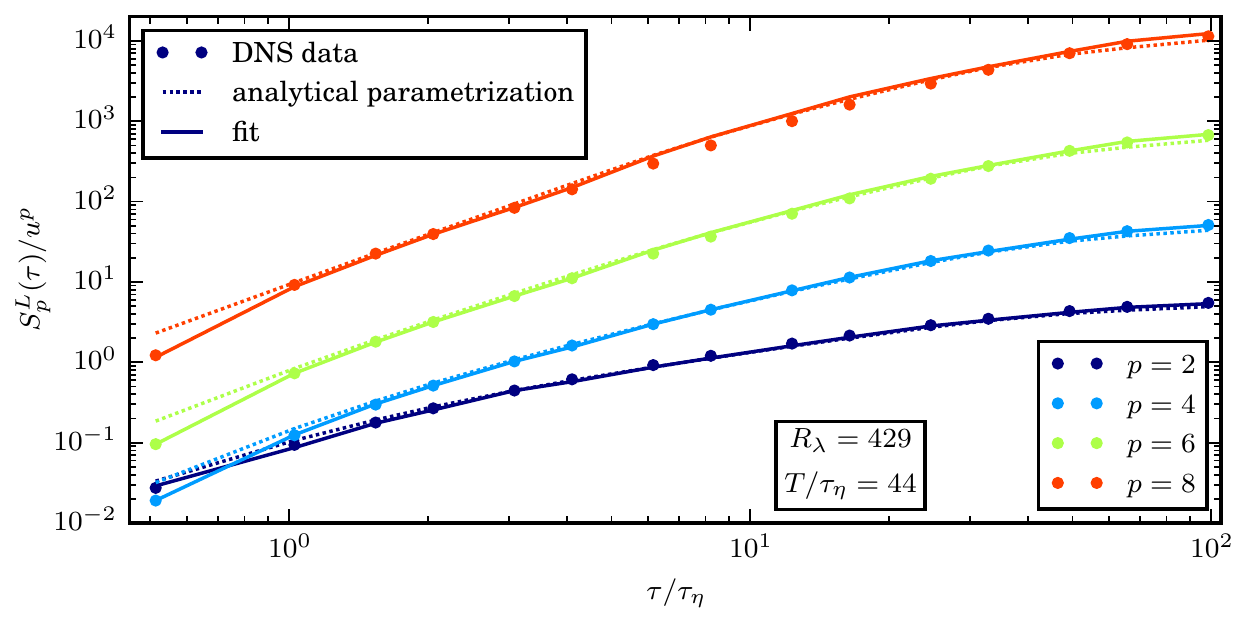}
\caption{
Comparison of Lagrangian structure functions for $R_\lambda = 429$: DNS data (filled circles) and results of applying the bridging relation with $\sigma(\tau)$ and $\mu(\tau)$ fitted to minimize \eref{eq:moment_distance} (continuous lines), as well as using the analytic expressions \eref{eq:fittedsigma} and \eref{eq:fittedmu}, respectively, for $\sigma(\tau)$ and $\mu(\tau)$ (dotted lines).
}
\label{fig:structurefunctions}
\end{figure}

\label{sisec:fittingprocedure}

The bridging relation \eref{eq:magnitudebridging} from the main text requires explicit knowledge of $\sigma(\tau)$ and $\mu(\tau)$.
While our theoretical arguments provide estimates for very short and very long times, data from direct numerical simulations is needed to obtain information for the full range of time scales.
The fitting procedure to obtain $\sigma(\tau)$ and $\mu(\tau)$ is explained in detail in the following.

From our numerical data, we obtain the PDFs of Lagrangian velocity increment magnitudes $f^L(v; \tau)$.
Our subsequent fitting is based on structure functions of these PDFs, which serve as a reference; the $p$th-order structure function is defined as \OK
\begin{equation}
    S^L_p(\tau) = \int \mathrm{d}v \, v^p f^L(v; \tau) \, .
    \label{eq:structure_function_definition}
\end{equation}
For the fitting, even-order structure functions up to order eight are considered.
To evaluate the structure functions from the right-hand side of the bridging relation \eref{eq:magnitudebridging}, we furthermore obtain the Eulerian velocity increment magnitude PDFs $f^E(v; R)$ directly from velocity field data.
Both Lagrangian and Eulerian data are averaged over the full simulation time $t_1-t_0$.

For fixed $\tau$ and given initial values of $\sigma$ and $\mu$,  model PDFs $\tilde{f}^L(v; \tau)$ are evaluated from \eref{eq:magnitudebridging}.
Because the bridging relation mixes Eulerian velocity increments for different scales depending on the velocity increment magnitude, the model PDFs need to be explicitly normalized.
The normalization factor $N(\tau)$, computed at the end of the fitting procedure, is shown in figure \ref{fig:Ntau}.
For time lags below one $\tau_{\eta}$, for which the dependence on the velocity increment is most pronounced, $N(\tau)$ deviates significantly from unity.
As expected, this deviation decreases with increasing time lag.

After computing the model PDFs, the corresponding model structure functions $\tilde S_p^L(\tau)$ are computed. These structure functions are then compared to the $S_p^L(\tau)$ obtained directly from the Lagrangian data. By minimizing the logarithm of \OK
\begin{equation}
    \mathcal{D}(\tau) \equiv \sum_{p\in\{2, 4, 6, 8\}}
        \frac{1}{p}\left(\frac{\left|\tilde{S}_p^L(\tau) - S_p^L(\tau)\right|}{
                               \min(\tilde{S}_p^L(\tau), S_p^L(\tau))}
                         \right)^{2/p}.
    \label{eq:moment_distance}
\end{equation}
we obtain best fits for $\sigma$ and $\mu$.
For the results presented, the ``Nelder-Mead'' algorithm \cite{nelder65tcj,wright96na} (through the \texttt{minimize} method available in the \texttt{scipy.optimize}~\cite{scipy} package) is used.
Repeating this procedure for all samples of $\tau$ under consideration, we obtain estimates for $\sigma(\tau)$ and $\mu(\tau)$, which then inspired our analytical parameterizations \eref{eq:fittedsigma} and \eref{eq:fittedmu} presented in the main text.
For the analytical $\mu(\tau)$ we combine the short-time prediction of $\tau/2$ with an exponential drop-off for larger times to take into account the decreasing correlation with the velocity increment.
It has to be pointed out that the precise determination of the drop-off of $\mu(\tau)$ becomes technically more difficult at long times where its influence is almost negligible.
For $\sigma(\tau)$, the data indicates three regimes, which we capture analytically by an extension of Batchelor interpolation \cite{she10ams}.
We then iterate this procedure taking our analytical parameterizations as improved initial guesses until convergence is reached.
In figure \ref{fig:fitsigma}, a direct comparison between the results of the fitting procedure for $\sigma(\tau)$ and the corresponding empirical formula \eref{eq:fittedsigma} for various Reynolds numbers shows good agreement.
The resulting structure functions from the DNS data, along with the ones obtained from the fitting procedure as well as from the analytical parameterizations, are shown in figure \ref{fig:structurefunctions}.

\section*{References}
% Bibliography


\begin{thebibliography}{10}

\bibitem{sreenivasan97arf}
K.~R. Sreenivasan and R.~A. Antonia.
\newblock The phenomenology of small-scale turbulence.
\newblock {\em Annu. Rev. Fluid Mech.}, 29(1):435--472, 1997.

\bibitem{ishihara07jfm}
T.~Ishihara, Y.~Kaneda, M.~Yokokawa, K.~Itakura, and A.~Uno.
\newblock Small-scale statistics in high-resolution direct numerical simulation
  of turbulence: Reynolds number dependence of one-point velocity gradient
  statistics.
\newblock {\em J. Fluid Mech.}, 592:335--366, 2007.

\bibitem{schumacher14pnas}
J.~Schumacher, J.~D. Scheel, D.~Krasnov, D.~A. Donzis, V.~Yakhot, and K.~R.
  Sreenivasan.
\newblock Small-scale universality in fluid turbulence.
\newblock {\em Proc. Natl. Acad. Sci. U.S.A.}, 111(30):10961--10965, 2014.

\bibitem{douady91prl}
S.~Douady, Y.~Couder, and M.~E. Brachet.
\newblock Direct observation of the intermittency of intense vorticity
  filaments in turbulence.
\newblock {\em Phys. Rev. Lett.}, 67:983--986, 1991.

\bibitem{cadot95pof}
O.~Cadot, S.~Douady, and Y.~Couder.
\newblock Characterization of the low-pressure filaments in a three-dimensional
  turbulent shear flow.
\newblock {\em Phys. Fluids}, 7(3):630--646, 1995.

\bibitem{ganapathisubramani08jfm}
B.~Ganapathisubramani, K.~Lakshminarasimhan, and N.~T. Clemens.
\newblock Investigation of three-dimensional structure of fine scales in a
  turbulent jet by using cinematographic stereoscopic particle image
  velocimetry.
\newblock {\em J. Fluid Mech.}, 598:141--175, 2008.

\bibitem{elsinga10jfm}
G.~E. Elsinga and I.~Marusic.
\newblock Universal aspects of small-scale motions in turbulence.
\newblock {\em J. Fluid Mech.}, 662:514--539, 2010.

\bibitem{worth11pre}
N.~A. Worth and T.~B. Nickels.
\newblock Time-resolved volumetric measurement of fine-scale coherent
  structures in turbulence.
\newblock {\em Phys. Rev. E}, 84:025301, 2011.

\bibitem{ashurst87pof}
Wm.~T. Ashurst, A.~R. Kerstein, R.~M. Kerr, and C.~H. Gibson.
\newblock Alignment of vorticity and scalar gradient with strain rate in
  simulated {N}avier–{S}tokes turbulence.
\newblock {\em Phys. Fluids}, 30(8):2343--2353, 1987.

\bibitem{she90nat}
Z.-S. {She}, E.~{Jackson}, and S.~A. {Orszag}.
\newblock {Intermittent vortex structures in homogeneous isotropic turbulence}.
\newblock {\em Nature}, 344:226--228, 1990.

\bibitem{vincent91jfm}
A.~Vincent and M.~Meneguzzi.
\newblock The spatial structure and statistical properties of homogeneous
  turbulence.
\newblock {\em J. Fluid Mech.}, 225:1--20, 1991.

\bibitem{jimenez93jfm}
J.~Jim{\'e}nez, A.~A. Wray, P.~G. Saffman, and R.~S. Rogallo.
\newblock The structure of intense vorticity in isotropic turbulence.
\newblock {\em J. Fluid Mech.}, 255:65--90, 1993.

\bibitem{vincent94jfm}
A.~Vincent and M.~Meneguzzi.
\newblock The dynamics of vorticity tubes in homogeneous turbulence.
\newblock {\em J. Fluid Mech.}, 258:245--254, 1994.

\bibitem{moisy04jfm}
F.~Moisy and J.~Jim{\'e}nez.
\newblock Geometry and clustering of intense structures in isotropic
  turbulence.
\newblock {\em J. Fluid Mech.}, 513:111--133, 2004.

\bibitem{leung12jfm}
T.~Leung, N.~Swaminathan, and P.~A. Davidson.
\newblock Geometry and interaction of structures in homogeneous isotropic
  turbulence.
\newblock {\em J. Fluid Mech.}, 710:453--481, 2012.

\bibitem{yeung15pnas}
P.~K. Yeung, X.~M. Zhai, and Katepalli~R. Sreenivasan.
\newblock Extreme events in computational turbulence.
\newblock {\em Proc. Natl. Acad. Sci. U.S.A.}, 112(41):12633--12638, 2015.

\bibitem{mordant01prl}
N.~Mordant, P.~Metz, O.~Michel, and J.-F. Pinton.
\newblock Measurement of {L}agrangian velocity in fully developed turbulence.
\newblock {\em Phys. Rev. Lett.}, 87:214501, 2001.

\bibitem{laporta01nat}
A.~{La Porta}, G.~A. Voth, A.~M. Crawford, J.~Alexander, and E.~Bodenschatz.
\newblock Fluid particle accelerations in fully developed turbulence.
\newblock {\em Nature}, 409:1017--1019, 2001.

\bibitem{mordant04phd}
N.~Mordant, A.M. Crawford, and E.~Bodenschatz.
\newblock Experimental {L}agrangian acceleration probability density function
  measurement.
\newblock {\em Physica D}, 193:245 -- 251, 2004.
\newblock Anomalous distributions, nonlinear dynamics, and nonextensivity.

\bibitem{biferale06jot}
L.~Biferale, G.~Boffetta, A.~Celani, A.~Lanotte, and F.~Toschi.
\newblock Lagrangian statistics in fully developed turbulence.
\newblock {\em J. Turbul.}, 7:N6, 2006.

\bibitem{kamps09pre}
O.~Kamps, R.~Friedrich, and R.~Grauer.
\newblock {Exact relation between Eulerian and Lagrangian velocity increment
  statistics}.
\newblock {\em Phys. Rev. E}, 79:066301, 2009.

\bibitem{homann09njp}
H.~Homann, O.~Kamps, R.~Friedrich, and R.~Grauer.
\newblock {Bridging from Eulerian to Lagrangian statistics in 3D hydro- and
  magnetohydrodynamic turbulent flows}.
\newblock {\em New J. Phys.}, 11(7):073020, 2009.

\bibitem{corrsin59agp}
S.~Corrsin.
\newblock Progress report on some turbulent diffusion research.
\newblock volume~6 of {\em Advances in Geophysics}, pages 161 -- 164. Elsevier,
  1959.

\bibitem{kraichnan64pof}
R.~H. Kraichnan.
\newblock {Kolmogorov's hypothesis and Eulerian turbulence theory}.
\newblock {\em Phys. Fluids}, 7:1723--1734, 1964.

\bibitem{tennekes75jfm}
H.~Tennekes.
\newblock {Eulerian and Lagrangian time microscales in isotropic turbulence}.
\newblock {\em J.~Fluid Mech.}, 67:561--567, 1975.

\bibitem{kolmogorov91prs}
A.~N. Kolmogorov.
\newblock The local structure of turbulence in incompressible viscous fluid for
  very large {R}eynolds numbers.
\newblock {\em Proc. Roy. Soc. London Ser. A}, 434(1890):9--13, 1991.

\bibitem{monin07book}
A.S. Monin and A.M. Yaglom.
\newblock {\em Statistical Fluid Mechanics: Mechanics of Turbulence},
  volume~II.
\newblock Dover Publications, 2007.

\bibitem{yakhot08arx}
V.~{Yakhot}.
\newblock {Lagrangian Structure Functions in Turbulence: Scaling Exponents and
  Universality}.
\newblock {\em ArXiv e-prints}, page 0810.2955, 2008.

\bibitem{boffetta02pre}
G.~Boffetta, F.~De~Lillo, and S.~Musacchio.
\newblock Lagrangian statistics and temporal intermittency in a shell model of
  turbulence.
\newblock {\em Phys. Rev. E}, 66:066307, 2002.

\bibitem{biferale04prl}
L.~Biferale, G.~Boffetta, A.~Celani, B.~J. Devenish, A.~Lanotte, and F.~Toschi.
\newblock Multifractal statistics of {Lagrangian} velocity and acceleration in
  turbulence.
\newblock {\em Phys. Rev. Lett.}, 93:064502, 2004.

\bibitem{borgas93ptr}
M.~S. Borgas.
\newblock {The Multifractal {Lagrangian} Nature of Turbulence}.
\newblock {\em Philos. T. R. Soc. A}, 342(1665):379--411, 1993.

\bibitem{chevillard12crp}
L.~Chevillard, B.~Castaing, A.~Arneodo, E.~L{\'{e}}v{\^{e}}que, J.-F. Pinton,
  and S.~G. Roux.
\newblock {A phenomenological theory of Eulerian and Lagrangian velocity
  fluctuations in turbulent flows}.
\newblock {\em C. R. Phys.}, 13(9-10):899--928, 2012.

\bibitem{corrsin63jas}
S.~Corrsin.
\newblock Estimates of the relations between {E}ulerian and {L}agrangian scales
  in large {R}eynolds number turbulence.
\newblock {\em J. Atmos. Sci.}, 20(2):115--119, 1963.

\bibitem{lanotte13jot}
A.S. Lanotte, L.~Biferale, G.~Boffetta, and F.~Toschi.
\newblock A new assessment of the second-order moment of {L}agrangian velocity
  increments in turbulence.
\newblock {\em J. Turbul.}, 14(7):34--48, 2013.

\bibitem{sawford11pof}
B.~L. Sawford and P.~K. Yeung.
\newblock Kolmogorov similarity scaling for one-particle {L}agrangian
  statistics.
\newblock {\em Phys. Fluids}, 23(9):091704, 2011.

\bibitem{busse07pop}
A.~Busse, W.-C. M\"uller, H.~Homann, and R.~Grauer.
\newblock Statistics of passive tracers in three-dimensional
  magnetohydrodynamic turbulence.
\newblock {\em Phys. Plasmas}, 14(12):122303, 2007.

\bibitem{hou07jcp}
T.~Y. Hou and R.~Li.
\newblock Computing nearly singular solutions using pseudo-spectral methods.
\newblock {\em J. Comput. Phys.}, 226:379--397, 2007.

\bibitem{shu88jcp}
C.-W. Shu and S.~Osher.
\newblock Efficient implementation of essentially non-oscillatory
  shock-capturing schemes.
\newblock {\em J. Comput. Phys.}, 77(2):439--471, 1988.

\bibitem{Lundgren03arb}
T.~S. Lundgren.
\newblock Linearly forced isotropic turbulence.
\newblock {\em Annual Research Briefs (Center for Turbulence Research,
  Stanford)}, pages 461--473, 2003.

\bibitem{atkinson89book}
K.~E. Atkinson.
\newblock {\em An Introduction to Numerical Analysis}.
\newblock John Wiley \& Sons, Inc., 2nd edition, 1989.

\bibitem{nelder65tcj}
J.~A. Nelder and R.~Mead.
\newblock A simplex method for function minimization.
\newblock {\em The Computer Journal}, 7:308--313, 1965.

\bibitem{wright96na}
M.~H. Wright.
\newblock Direct search methods: Once scorned, now respectable.
\newblock {\em Numerical analysis 1995: Proceedings of the 1995 Dundee Biennial
  Conference in Numerical Analysis}, pages 191--208, 1996.
\newblock D. F. Griffiths and G. A. Watson (eds.).

\bibitem{scipy}
E.~Jones, T.~Oliphant, P.~Peterson, et~al.
\newblock {SciPy}: {Open} source scientific tools for {Python}, 2001--.
\newblock [Online; accessed 2017-04-10].

\bibitem{she10ams}
Z.-S. She, X.~Chen, Y.~Wu, and F.~Hussain.
\newblock New perspective in statistical modeling of wall-bounded turbulence.
\newblock {\em Acta Mech. Sinica}, 26(6):847--861, 2010.

\end{thebibliography}
\end{document}